\documentclass[a4paper,fleqn]{cas-dc}

\usepackage[numbers]{natbib}
\usepackage{graphicx}
\usepackage{multirow}
\usepackage{xcolor}
\usepackage{flushend}
\def\tsc#1{\csdef{#1}{\textsc{\lowercase{#1}}\xspace}}
\tsc{WGM}
\tsc{QE}
\tsc{EP}
\tsc{PMS}
\tsc{BEC}
\tsc{DE}

\begin{document}
\let\WriteBookmarks\relax
\shorttitle{DemirciPro tools}
\shortauthors{O. Cakir et~al.}
\title [mode = title]{DemirciPro's tools for completing the Linac: Ion source and LEBT line}                      
\tnotemark[1]



\author[1]{Orhan Cakir}
\address[1]{Ankara University, Department of Physics, Ankara, Turkey}
\author[2]{Emre Celebi}[style=turkish]
\address[2]{Bogazici University, Department of Physics, Istanbul, Turkey}
\author[3]{Hakan Cetinkaya}
\address[3]{Dumlupinar University, Department of Physics, Kutahya, Turkey}
\author[4] {Hilal Kolenoglu}
\address[4]{Eskisehir Teknik University, Department of Physics, Eskisehir, Turkey}

\author[5] {Gorkem Turemen}
\address[5]{Turkish Atomic Energy Authority, Division of Proton Accelerator, Department of Radiation and Accelerator Technologies, Ankara, Turkey}
\author[6] {Zekeriya Uysal}
\address[6]{Gaziantep University, Department of Engineering Physics, Gaziantep, Turkey}
\author[7] {Gokhan Unel}
\address[7]{University of California at Irvine, Department of Physics and Astronomy, Irvine, USA}

\cortext[cor1]{Corresponding author}


\begin{abstract}
Demirci RFQ design software has been enlarged to perform other design tasks pertinent to the initial part of a linac. 
These are the design of an ion source, the design of a low energy beam transmission line including the realistic design of its magnets and  finally the design of a pepper pot emittance meter. 
This note presents these latest developments together with some examples from a linac being developed in KahveLab, Turkey. 

\end{abstract}


\maketitle
\section{Introduction}
\label{sec:intro}

Since the first Radio Frequency Quadrupole (RFQ) accelerator concept was invented and demonstrated in Protvino, Russia, in 1972 \cite{Kapchinskii1969ALI} and later improved at Los Alamos in 1980s \cite{Biscari1985ComputerPA} many RFQs have been built worldwide for a variety of valuable applications in research, industry and medicine. However since the early days, the number of computer programs dedicated to RFQ design has been small. Examples of commonly used, (and commercial)  design programs can be given as: Lidos \cite{Bondarev:2001in}, Parmteq \cite{Crandall1988PARMTEQABD} and RFQGEN \cite{RFQGEN}. Some of such software like Benelos \cite{BENELOS} are no longer available. Some of these programs have also the capability of designing other sections of the beamline in addition to the RFQ simulations. For example Parmteq can also design the Low Energy Beam Transport (LEBT) Line \cite{Jang2006, ParmteqManual}, but does not contain an ion source section. Toutatis can only do the beam dynamics of the RFQ and it can also help with the design the LEBT and pepper pot meter via the associated Tracewin program \cite{Pande_2015, Uriot2015}). Since 2013, another design software, with a radically new RFQ design method, has been actively developed: Demirci. On  top of the classical two term potential initial design and 8 term potential beam dynamics calculations, it has the unique property of being a fully graphical user interface (GUI) based. In this novel approach, to test a new design idea, the designer is expected to simply drag a few key points to their new positions.

\begin{table}
 \caption{DemirciPro vs other similar software}
 \centering
 \begin{tabular}{|r|c|c|c|c|c|}
\hline
Code     & IS   & LEBT & PPM & RFQ  & RBD \\ \hline
Toutatis &  -  &  *   &  *        &    -       &         +    \\ \hline
ParmteqM &  -  &  +   &   -       &    +       &         +    \\ \hline
Lidos-RFQ&  -  &  -   &   -       &    +       &         +    \\ \hline
DemirciPRO& +* &  +   &   +       &    +       &         +    \\ \hline
\end{tabular}\\
-: not available\\
*: with Tracewin support\\
+*: with IBSimu integration
\end{table}

The work on the second iteration of Demirci, called DemirciPRO, has started in 2017. This new version contains improvements and additions to make it a complete suite for light ion linear accelerator design. These additions are the design of an ion source, the design of an electromagnetic lens system to transport the ions to the RFQ cavity and finally a pepper pot emittance measurement setup. Therefore, from the ion source(IS), up to the RFQ exit, including the LEBT, pepper pot meter(PPM) \cite{Stockli2006}, RFQ design(RFQ) and RFQ beam dynamics(RBD) simulations can be performed by a single software. The comparison of DemirciPRO with other similar software in terms of their capabilities is presented in Table 1.
The remaining of this paper discusses the details of these new modules, their implementation and their validation. Some examples from an ongoing linac project will also be shown.
\section{IBSimu integration}
\label{sec:ibsimu}

\subsection{Overview of IBSimu}
IBSimu \cite{Kalvas2010} is an ion beam simulation library written in C++, mostly used for mimicking the extraction of ions from plasma, including the space-charge effects. It uses two or three-dimensional Cartesian or cylindrical coordinates and has a nonlinear Finite Difference Method (FDM) Poisson's equation solver for electrostatic fields. 

Electrode definitions, plasma temperature, current density, boundary potentials, electrode voltages, plasma depth, mesh size are some of the required input parameters for IBSimu. 
Functions importing DXF, and STL files can be accepted as an electrode definition where these functions are Boolean analytic definitions of constraints on x,y and z-axis. IBSimu tracks the charged particles under the electric and magnetic fields, simulates positive and negative plasma extraction, and also calculates the space charge and electric potential distributions. In addition, the magnetic field can be imported into the simulations \cite{Toivanen2013}.

The simulation starts by constructing the volume with imported electrode files and defined mesh sizes. Next step is to render a discrete version of this volume with a rectangular mesh. The potential distribution is solved using Poisson's equation after considering the ion source  (IS) geometry, electrode potentials and the boundary conditions. 
Firstly the particle trajectories are calculated; secondly the  space charge effect on the beam is propagated to the mesh nodes and it is taken into account while solving the electric potentials iteratively. This chain is repeated until the solution converges\cite{Toivanen2013}. 

\subsection{DemirciPro IS Module}
\begin{figure}
\centering
\includegraphics[width=0.99\linewidth]{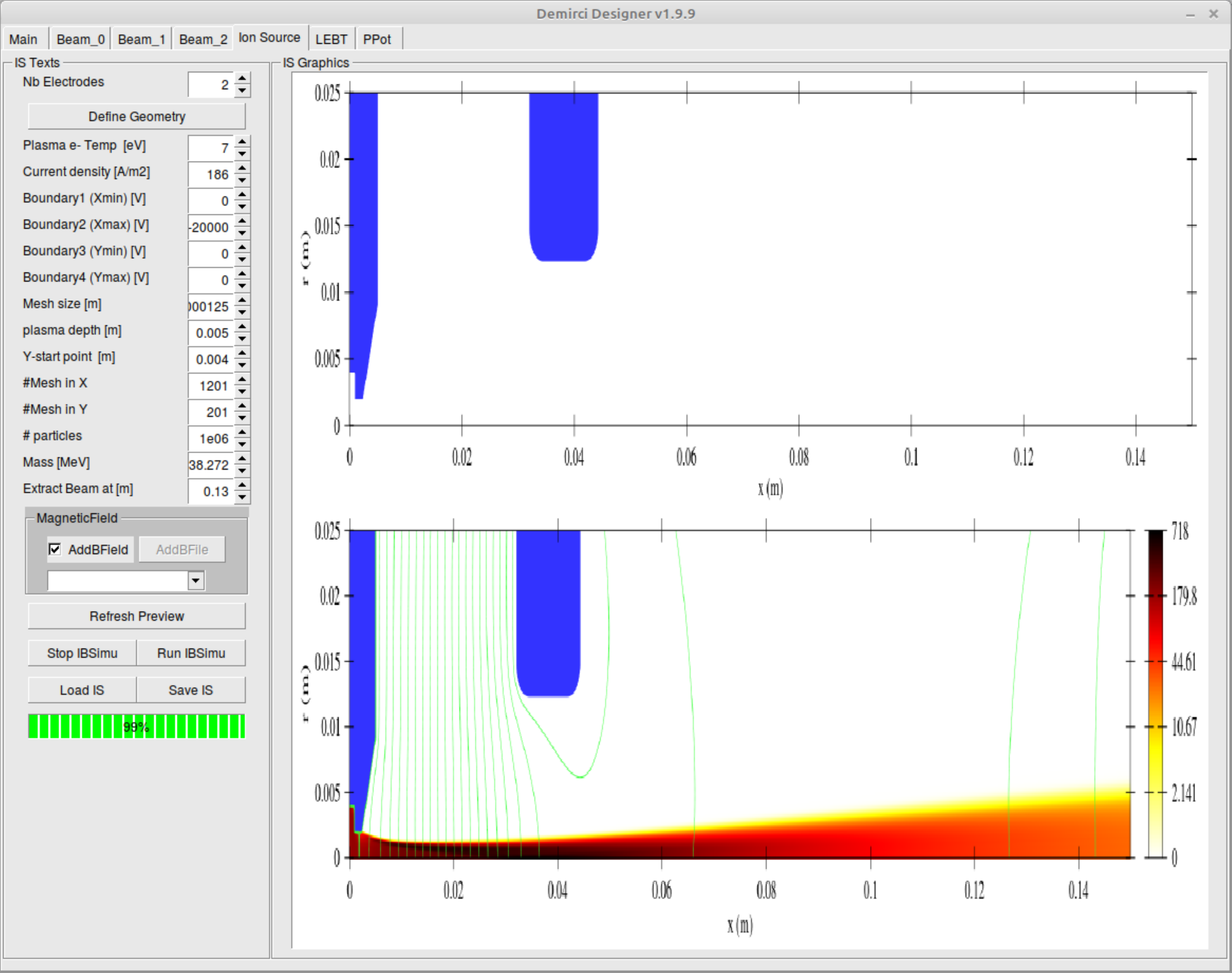}
\caption{Ibsimu Interface}
\label{fig:ibsimuint}
\end{figure}

A user-friendly graphical interface (GUI) has been developed and integrated into the DemirciPro ion Linac design platform. This new interface, written in C++ using ROOT \cite{Brun:1997pa} and IBSimu \cite{Kalvas2010} libraries, allows an ion source design and  its beam simulation in cylindrical-coordinates. The GUI, seen in Figure \ref{fig:ibsimuint}, offers a set of configuration and monitoring tools as well as the related outputs in DST format to simplify the designer's workload. Cylindrical coordinates have been selected in simulation since most of the ion sources have cylindrical symmetry and it gives an advantage for CPU time and memory consumption in providing solutions to Poisson's equation. 

IS interface has the capability of simulating both positive and negative ions. The DemirciPro ion beam section uses a very basic  user interface to set the values to define the IS setup parameters. These parameters can be saved and reloaded as required. This simple setup allows the user to concentrate on the simulation and to optimize the parameter set efficiently. It also offers design practicality in IS configuration: The user is warned via a pop-up window if the parameters are out of the defined boundary conditions. The ion charge and mass can also be set to the desired values using DemirciPRO internal parameters setup panel. 
The boundary voltages are adjusted using the left side of the panel. The first boundary is the ion insertion point. The Dirichlet boundary condition is applied for the negative ion source, while the Neumann boundary condition for the positive one. These are provided as default as indicated in the IBSimu reference manual \cite{IBSimuManual}. 

\begin{figure}
\centering
\includegraphics[width=0.99\linewidth]{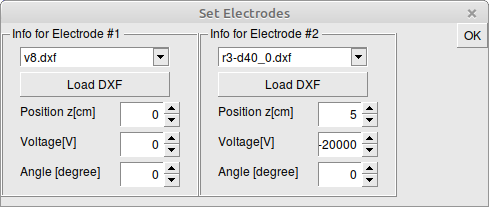}
\caption{ Electrode selection window}
\label{fig:electrodsel}
\end{figure}

Electrode CAD drawings in DXF format can be imported into the available geometry pool. These imported electrodes are then picked from a drop-down list box in the electrode selection window shown in Figure \ref{fig:electrodsel}. Each electrode can be positioned along the beam axis (z position), it can be rotated (in degrees) around the radial axis and its electric potential can be set in Volts. This GUI offers the designer an opportunity to define every parameter of the electrodes in one step. CAD drawings are used primarily for complex geometries which would be difficult to describe by analytical functions.

The preview of selected electrode geometries and their positions in cylindrical coordinates can be seen in the upper right section of the Figure \ref{fig:ibsimuint}. The lower section shows the simulation results that are plotted as a function of extracted beam current density. The green progress bar indicates the completed percentage of the running simulation.
An external magnetic field file can also be defined and imported in DemirciPro. 
Such a field map can be incorporated into the simulation by selecting a magnetic field map file using the drop-down list in the dark gray section of Figure \ref{fig:ibsimuint}.
 There may be a residual magnetic field in the extraction zone of the ion source due to the magnetic field applied around the plasma chamber, the selected map file can be used to represent that residual field, or alternatively, the designer may simply want to apply a magnetic field in the extraction region of the ion source for testing purposes. The magnetic field map file is basically a 4 column tab-separated text file,  defining the z~[m], r~[m], Bz~[T], and Br~[T], representing the position along the beamline, radial position, axial magnetic field strength and radial magnetic field strength, respectively. 

IBSimu can export the beam data at the desired position in an output file. This file would be used as an input to a design software such as Travel/Path Manager~\cite{TRAVELManual}, and the results would finally be represented with a graphics tool such as PlotWin~\cite{PlotwinManual}. DemirciPRO shortens this procedure by providing the necessary functions and a single user interface to these tasks. The user may specify an ion beam extraction point along the z-axis at which the beam data, i.e. particle information, will be exported in both as DST and TXT file formats. The TXT format is a tab-separated text file that contains the 6D particle information, whereas DST format is a binary file that contains the same information. Some commonly used tools in the field, such as PlotWin,  use only DST format input files.

\begin{figure}
\centering
\includegraphics[width=0.99\linewidth]{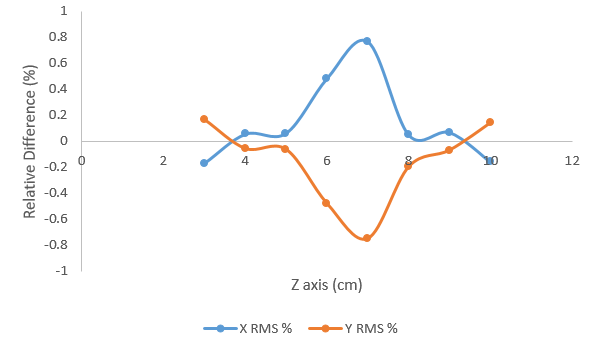}
\caption{IBSimu and DemirciPro  comparison}
\label{fig:IBSvsDemirci}
\end{figure}

 To ensure the correct operation of DemirciPro, it was compared to the standalone version of IBSimu using the same electrode and parameter configuration. Then, the same extraction geometry was implemented in both
 Travel/Path Manager and DemirciPro to compare the beam behaviour under the conditions without the space charge effects.  Figure \ref{fig:IBSvsDemirci} shows that the relative difference for the beam envelope along the beamline is less than 0.7 percent. The Relative Difference (RD) was calculated with respect to standalone IBSimu values as:
\begin{equation}\label{eq:rd}
\texttt{RD}(\%) = \frac{(RMS_{Demirci}- RMS_{IBSimu})} {RMS_{IBSimu}}\times 100 \qquad .
\end{equation}
 The simulations were performed for 8 data points in the z axis with about 20'000 particles. 
  The DST writer in DemirciPRO is based on \texttt{export\_path\_manager\_data }function in IBSimu. As it uses the same algorithm to calculate the x and y values of the particles, the observed difference is due mainly  to the random number generator and output formats. It is also known that there are small differences in RMS values between the exported Path Manager outputs in cylindrical coordinates due to the nature of randomized phase and azimuthal angle. 

\section{LEBT Design}
\label{sec:lebt}

Low Energy Beam Transport (LEBT) line is used to transfer ion beam from ion source to the next stage of the linac, which usually is the Radio Frequency Quadrupole (RFQ). It is used to match the ion beam in diameter and in emittance to the acceptance values of the RFQ. Usually, some beam diagnostic elements such as steerer magnets and emittance measurement tools are also placed along the LEBT line. At least two electromagnetic lenses are required to optimize the beam in the LEBT line for higher transport efficiencies. It basically amounts to matching the Twiss parameters ($\alpha_x, \beta_x$) to their radially symmetric vertical counter parts  ($\alpha_y, \beta_y$) which leads to have the RMS emittance of the beam 9-fold smaller than the RFQ acceptance resulting in about 90\% beam acceptance~\cite{Keller1999}.

Although there are two types of LEBT lines, electrostatic and magnetic, there are no differences between these in terms of particle optics. Electrostatic LEBT lines consist of several electrodes and their simulation can be made by IBSimu \cite{Kalvas2010}. Magnetic LEBT lines consist of solenoid, quadrupole, dipole and steerer magnets depending on the requirements of the facility \cite{Keller1999}. Two solenoid magnet systems or Einzel lenses are more commonly used in magnetic LEBT setups, especially in high current beam lines, compared to quadrupole magnets \cite{Prost2016,Chauvin2011}. 

\begin{figure}
\centering
\includegraphics[width=1.5\linewidth, angle=270]{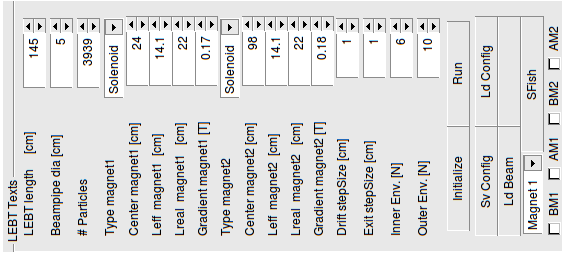}
\caption{DemirciPro LEBT section input parameters}
\label{fig:Lebtinput}
\end{figure}

DemirciPro has a LEBT design section which contains a magnetic LEBT line simulation code and a ROOT \cite{Brun:1997pa} based GUI to interface it. This LEBT design section has two distinct ion beam input possibilities: it can either randomly create an ion beam based on the user defined Twiss parameters or it can read the beam written out by an ion source simulator in either DST or TXT formats. In the first case, the user can also specify the number of macro particles to be generated and tracked, whereas in the second case, all particles read from the beam definition file are used. 
Similar to the beam input file, all parameters of the LEBT configurations can also be saved and loaded by the designer using the appropriate buttons in the GUI. The currently allowed LEBT line magnet types can be specified by using a drop down menu: a solenoid, a quadrupole or a dipole. For each magnet, the physical (real) length, the effective magnetic length, the center position along the beamline and the magnetic field strength have to be specified. In order to correctly simulate the beam behaviour along the LEBT line, its total length and the diameter of the beam pipe are also  required input parameters. The input section of these configuration parameters together with the values from an ongoing LEBT line design can be found in Figure \ref{fig:Lebtinput}. 

\begin{figure}
\centering
\includegraphics[width=0.99\linewidth]{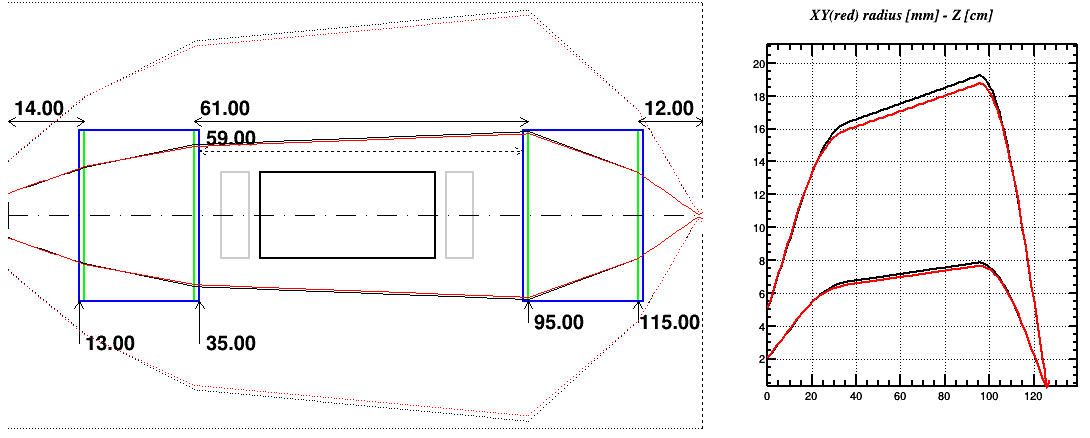}
\caption{Magnet locations and beam envelopes, see text for the explanations of various line colors and styles.}
\label{fig:beamenvelope}
\end{figure}

At the initialization of the LEBT line (using the appropriate button on the GUI), the location of the LEBT focusing magnets (physical size in blue), deflection magnets (in gray) and the measurement box (in black) are shown in the DemirciPRO GUI, for which an example is presented on the left side of Figure \ref{fig:beamenvelope}. The measurement box is a device which typically contains elements like a scintillator screen, a Faraday Cup and an emittance meter which will be covered in the next section.

The simulation section of the DemirciPRO software moves each particle from the beginning of the LEBT line towards the RFQ. At each step, each particle's position in the 6-D phase space is determined using the appropriate transfer matrices. Although the current version does not take the space charge effects into account, the improvements in the project continue, and this effect is planned to be incorporated in the next version of the software. In the LEBT calculations, only the drift space lengths and effective magnet lengths are used as the real magnet size is only provided for guiding the eye. For a realistic simulation it is obvious that the step size should be set as small as possible. However, depending on the number of tracked particles, this can consume some large CPU time. For example, a run with 20'000 particles and a step size of 1 mm took 5 seconds on an Intel-i5 CPU laptop.
For this reason, the step-size in drift areas can be determined by the designer, as well as the step-size in the region just before the RFQ. A LEBT example with 20'000 particles is shown in Figure \ref{fig:beamenvelope} .

Although all particles are tracked, it is hard to visualize their behaviour all at once. For this reason, two different beam envelopes, named inner and outer, are defined and plotted for x-z and y-z planes. The beam envelopes are defined as the root mean square (RMS) distributions, and the decision for selecting the quantile number for both inner and outer values is left to the LEBT designer. As an example, one can see from Figure \ref{fig:Lebtinput} that N=6 (inner) and N=10 (outer) RMS envelopes were selected. As usual, N=1 represents 39\% of the beam, N=6 correspond to about  95\% and finally N=10 to about 99\%. Having two different envelopes helps the designer to better estimate the beam behaviour, especially to see which portion of the beam halo would hit the beam pipe.  
In the left side of Figure \ref{fig:beamenvelope}, example beam halos are plotted, as solid (dashed) lines for inner (outer) and red (black) for x (y) axis along the LEBT line. In the same figure, still on the left side, blue boxes show physical lengths of the magnets while green ones represent the effective lengths, gray boxes show steerer magnets and finally the black box stands for the measurement box containing the diagnostic tools.  
The drawing is to scale along the beam axis (z), and also along the perpendicular axis for the beam pipe, shown by the horizontal dotted lines at the top and bottom of the drawing, and for beam envelopes. However the perpendicular (x or y) size of the magnets and of the measurement box is for representative purposes only.  In this particular example, the beam outer envelope approaches the beampipe of (radius=2.5~cm) at around 95~cm along the z axis. A more technical plot is also presented to the designer as can be seen on the right side of Figure \ref{fig:beamenvelope}: it contains the beam radius and z axis information for both beam envelopes.  The designer can zoom and check a particular section of this plot using standard ROOT canvas capabilities.

\begin{figure}
\centering
 \includegraphics[width=0.99\linewidth]{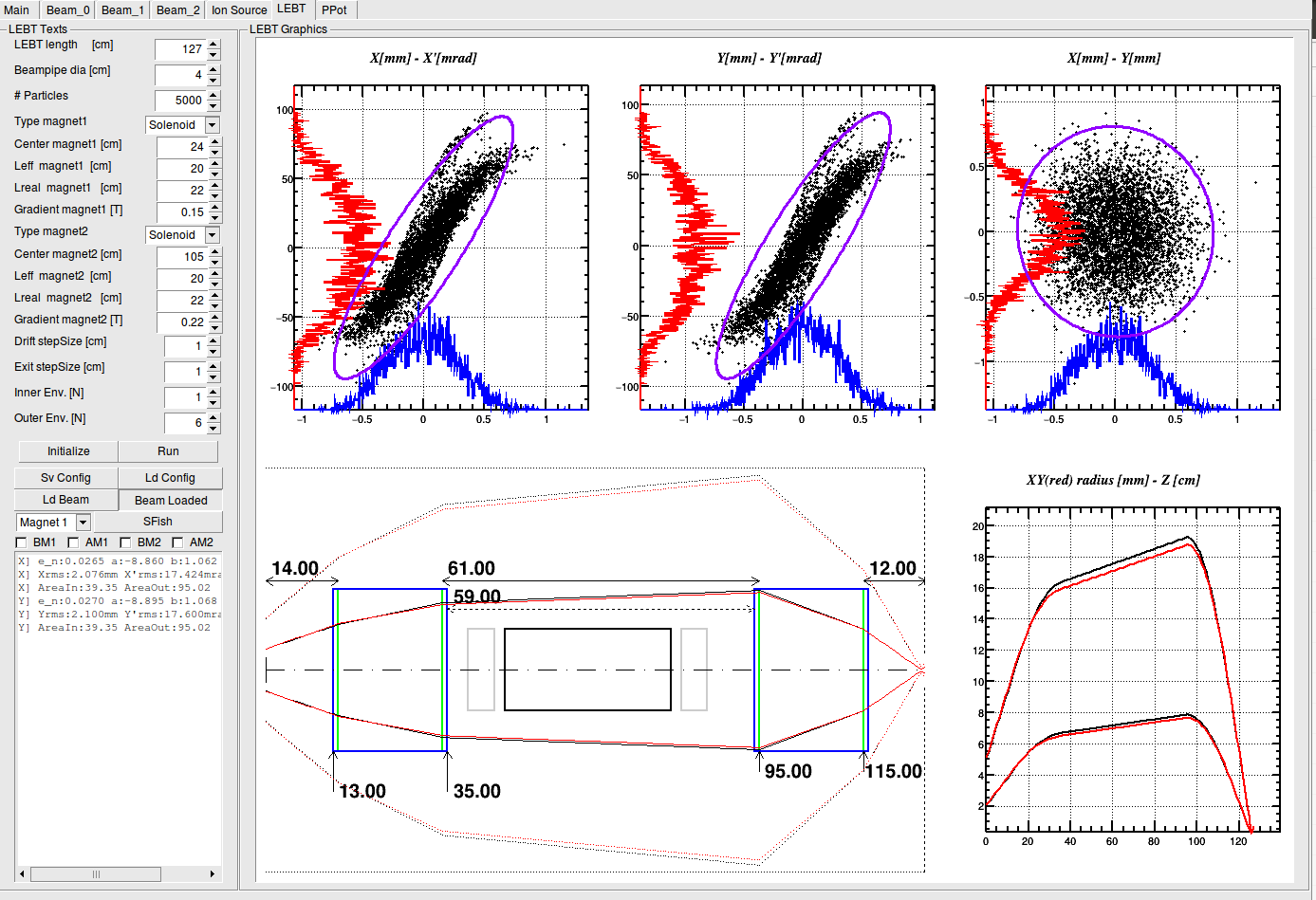}
\caption{DemirciPro Low Energy Beam Transport (LEBT) design module}
\label{fig:lebt}
\end{figure}

Transverse beam size and  emittances are plotted on the upper side of the output screen as they are calculated. An example of the whole LEBT tab of DemirciPRO can be seen in Figure \ref{fig:lebt}. It is also possible to save simulated particles as a beamfile in DST format along the LEBT line, before and after each magnet.  For speed considerations, this feature is made optional, saving beam files at different locations can be enabled from the left side of the GUI using standard checkboxes. These files can be used for further studies, such as comparison with other beamline simulation programs. The content of these files can be viewed either with DemirciPro itself, or with external programs such as Plotwin \cite{PlotwinManual}.

\begin{table}
\caption{Input beam properties and designed LEBT Line properties}
\centering
\begin{tabular}{|l|c|}
\hline
 Input Beam $\sigma_x$ (mm) & 1.325 \\
 \hline
 Input Beam $\sigma_y$ (mm) & 1.340 \\
 \hline
 Normalized RMS
 Emittance ($\pi$.mm.mrad) & 0.0259 \\
 \hline
 Sol 1 Bfield (kG) &1.5 \\
 \hline
 Sol 1 Effective Length (cm) &20 \\
 \hline
 Sol 1 center location (cm) &24 \\
 \hline
 Sol 1 Real Length (cm) & 22 \\
 \hline
 Sol 2 Bfield (kG) & 2.2 \\
 \hline
 Sol 2 Effective Length (cm) & 20 \\
 \hline
Sol 2 center location (cm) & 105 \\
 \hline
 Sol 2 Real Length (cm) & 22 \\
  \hline
 LEBT Length (cm) & 127 \\
   \hline
 \end{tabular}\\ 
 
 \label{tbl:rms}
 \end{table}

\begin{figure}
\centering
\includegraphics[width=1.0\linewidth]{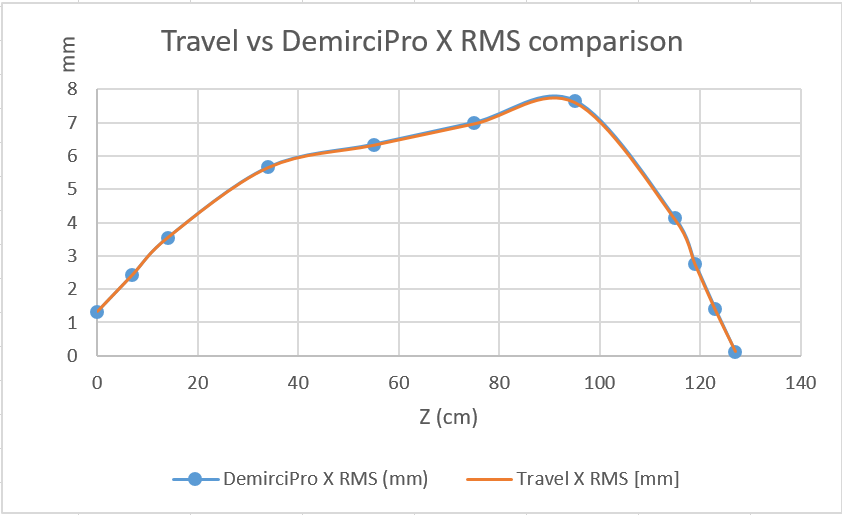}
\includegraphics[width=1.0\linewidth]{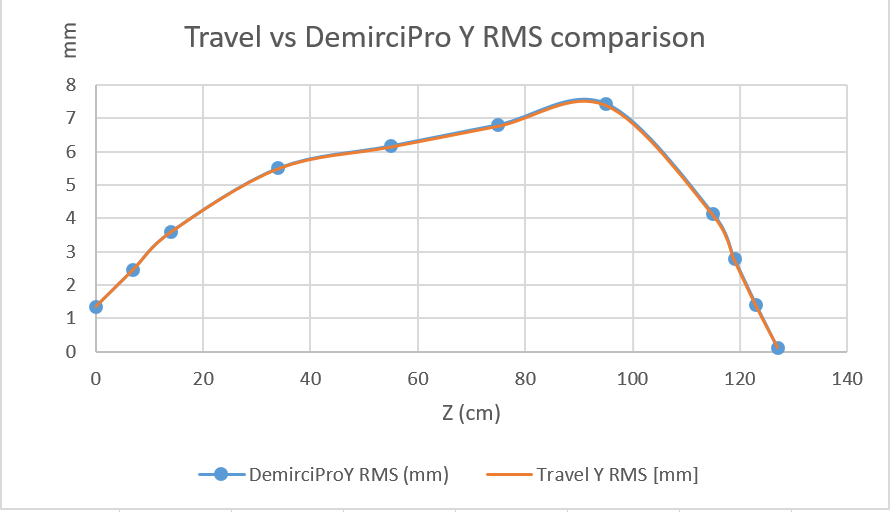}
\caption{RMS beam envelopes along the beam axis, (top: X direction, bottom: Y direction )}
\label{fig:BeamEnvelopeCmp}
\end{figure}

One of the software programs mostly used for magnetic LEBT line simulations is Travel/Path Manager \cite{TRAVELManual}. This software can simulate the beamline both with and without the space charge effects. Therefore, a comparison can be made between this well established software and DemirciPro to validate the latter, in the absence of the space charge effects. For the validation test, the simulation of a LEBT line, defined in Table \ref{tbl:rms}, was performed with 5000 particles described in a file written out by IBSimu. Drift and exit step sizes were set as 1 cm for this comparison. 
As seen in Table \ref{tbl:travelcompare}, and  in Figure \ref{fig:BeamEnvelopeCmp},
the computed RMS beam envelope values are compatible with each other at different locations along the beam axis, the largest difference being about 5\% at the beam waist.

 \begin{table}
 \caption{RMS beam envelope comparison. The first row corresponds to X direction and second row corresponds to Y direction at each location). RMS Pull is defined as $ \frac{RMS_{DemirciPro} - RMS_{Travel}}{RMS_{Travel}}\times 100$ .}
 \centering
 \begin{tabular}{|c|c|c|c|}
\hline
Location (cm) & Travel & DemirciPro & RMS Pull \% \\
\hline
\multirow{2}{*}{7}  & 2.432 & 2.432 & 0 \\
                    & 2.459 & 2.459 & 0 \\
\hline
\multirow{2}{*}{14} & 3.545 & 3.545 & 0 \\
                    & 3.585 & 3.585 & 0 \\
\hline
\multirow{2}{*}{34} & 5.651 & 5.657 & 0.103 \\
                    & 5.494 & 5.500 & 0.109 \\
\hline
\multirow{2}{*}{55} & 6.323 & 6.342 & 0.300 \\
                    & 6.149 & 6.167 & 0.305 \\
\hline
\multirow{2}{*}{75} & 6.965 & 6.996 & 0.451 \\
                    & 6.775 & 6.806 & 0.455 \\
 \hline
\multirow{2}{*}{95} & 7.609 & 7.653 & 0.576 \\
                    & 7.404 & 7.447 & 0.579 \\
 \hline
\multirow{2}{*}{115}& 4.075 & 4.126 & 1.232 \\
                    & 4.090 & 4.139 & 1.198 \\
 \hline
\multirow{2}{*}{119}& 2.716 & 2.765 & 1.807 \\
                    & 2.726 & 2.774 & 1.772 \\
 \hline
\multirow{2}{*}{127}& 0.116 & 0.122 & 5.372 \\
                    & 0.115 & 0.122 & 5.147 \\
 \hline
\end{tabular}
 \label{tbl:travelcompare}
 \end{table}
 
 \subsection{Poisson Superfish Integration}
 \label{sec:poisson}
\begin{figure}
 \centering
 \includegraphics[width=0.99\linewidth]{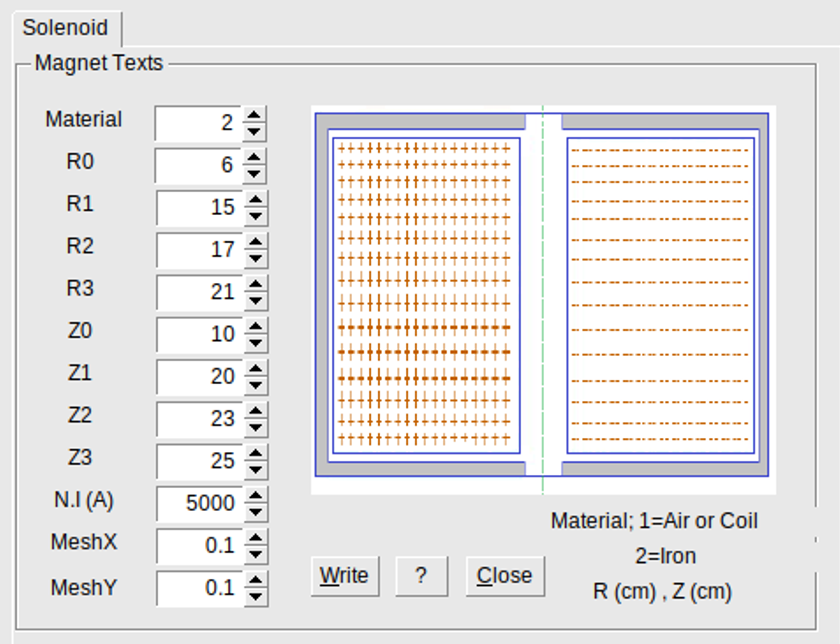}
 \caption{Parameters for geometric shape and numerical values for a solenoid magnet}
 \label{fig:sol-parameters}
 \end{figure}

 As part of the DemirciPro LEBT subsystem, it is possible to make a realistic magnet design by interfacing with the Poisson - Superfish software suite. This suite is a collection of programs for calculating static magnetic and electric fields and radio frequency electromagnetic fields in axial symmetrical cylindrical coordinates. It also contains graphic display tools and other similar codes to show the obtained results in a variety of ways.
 The interface to this suite is available through the SFish button (Figure \ref{fig:lebt}) on the left side of the LEBT design screen. The target magnet, as previously discussed in LEBT design part, can be either a quadrupole or a solenoid. When a magnet type is selected from the menu, the design parameters are displayed in the next window. An example for the solenoid design can be found in Fig.\ref{fig:sol-parameters}. 
 The magnet physical dimensions, the coil current and the core material type can also be configured.

\begin{figure}
 \centering
 \includegraphics[width=0.5\textwidth,height=0.4\textheight,keepaspectratio]{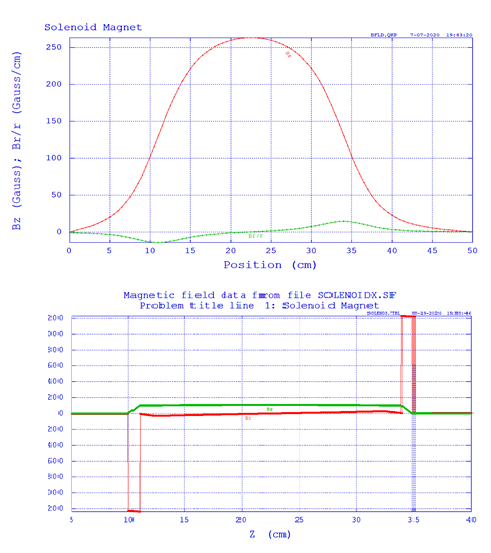}
 \caption{Solenoid magnetic field distribution (upper) through the direction of the beam (r = 0.42 cm) (lower) magnetic field distribution along r = 25.5 cm}
 \label{fig:selenoidmgfdist}
 \end{figure}

 The magnetic field inside a solenoid is designed to be approximately uniform; however, on the outside, it is weak and divergent.  In order to obtain a symmetrical field distribution, the left and right coils are considered to be identical but the current flow directions differ according to the design. Figure \ref{fig:selenoidmgfdist} shows the magnetic field distribution for the example design in Fig.\ref{fig:sol-parameters}. The red (green) curve represents the field in r (z) direction.
 
 \clearpage
 
\begin{figure}[!htbp]
 \centering
 \includegraphics[width=0.95\linewidth]{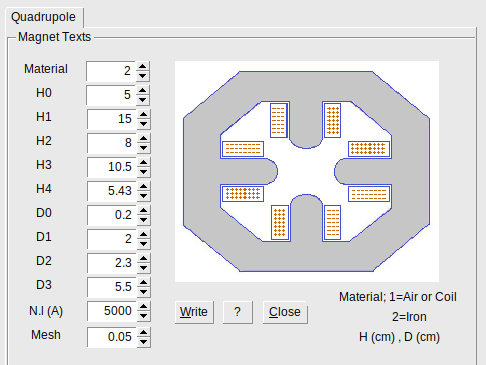}
 \caption{Geometric shape for parameters and numerical values for the skew quadrupole}
 \label{fig:sqmqGeoShape}
 \end{figure}

 The magnetic field in a quadrupole has two components showing a configuration with hyperbolic pole shapes in the perpendicular plane. The parameters of the geometry for quadrupole design are given in Fig \ref{fig:sqmqGeoShape}. On the user interface, the button with the question mark sign provides the skew quadrupole and normal quadrupole magnet definitions. Both configurations are written on files with sf extensions:  quadrupolnx.sf (normal) and quadrupolrx.sf (rotated or skew) . These files can be run with the SuperFish program to obtain the  magnetic field distributions.

  \begin{figure}[!htbp]
 \centering
 \includegraphics[width=0.95\linewidth]{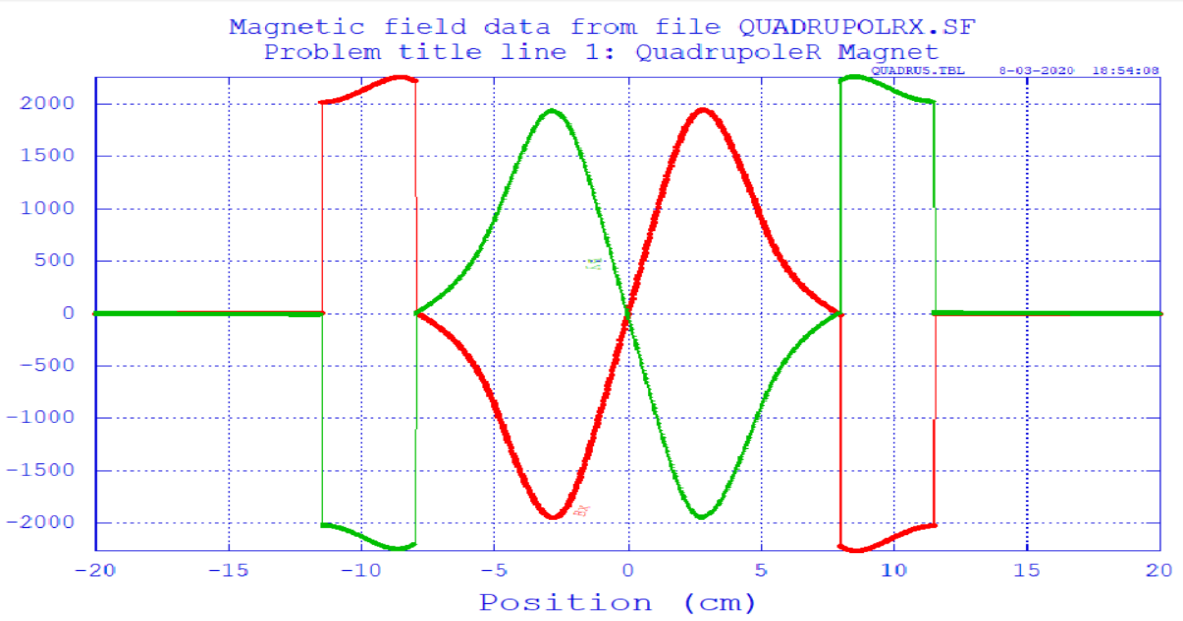}
 \caption{The distributions of magnetic field components for skew quadrupole}
 \label{fig:quadMGfieldDistNorm}
 \end{figure}

It is crucial to have a proper magnetic field distribution to correctly focus the ion beam. This implies a symmetrical field around the coils and zero field at the center as shown in Fig\ref{fig:quadMGfieldDistNorm}. For a beam of positively charged ions (or protons) directed towards the reader, such a quadrupole magnet focuses the beam in the vertical direction and defocuses it in the horizontal direction. When this quadrupole magnet is rotated by pi/2, then the beam is focused horizontally and defocused vertically.

The quadrupole or solenoid geometry created in DemirciPRO can be studied with the Poisson - Superfish software suite. The suit can be used directly on a computer running a Windows operating system or on a computer under Linux and OSX operating systems using recent versions of Wine compatibility layer software \cite{winehq}.
 After dividing the problem's geometry into a finite number of meshes, the solution file is created by solving the Poisson equation and the magnetic field lines are obtained. The  Poisson - Superfish software suite provides tools for plotting the field lines or for displaying the field value at a user specified point.

 \section{Measurement Box Design}
 \label{sec:pepper}
\begin{figure}
 \centering
 \includegraphics[width=0.95\linewidth]{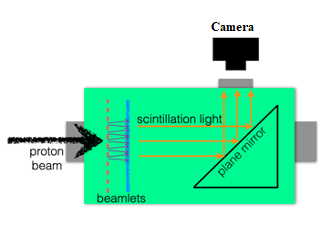}
 \caption{A schematic representation of the pepper pot emittance measurement setup}
 \label{fig:pepperpotschema}
 \end{figure}

 While building a beamline, one of the important goals is being able to characterize the beam properties. The typically measured quantities are the beam charge, the beam profile and the beam emittance. The beam charge is typically measured with a Faraday Cup, a destructive measurement, which could be purchased according to one's budget and expected beam properties. On the other hand, the setup for measuring the beam profile and emittance usually depends on the LEBT line and has to be specific to the designed beamline. DemirciPro offers an integrated section for designing a pepper pot and scintillator-based setup \cite{Stockli2006} for measuring these properties. This section uses the beam simulated in the LEBT section which needs to be executed beforehand. 
 In the emittance measurement setup, the incoming beam first hits the pepper pot plate which  allows a limited number of beamlets to pass. These beamlets then hit a scintillating screen, causing it to emit light which is captured by a camera. 
 The camera, installed over a peephole,  sees the scintillator through a plane mirror placed inside the measurement box. The picture obtained in the camera is later analyzed to deduce the beam parameters. A schematic representation of this setup is shown in Figure \ref{fig:pepperpotschema}. Although there are alternative methods for measuring the beam emittance, (such as the three-screen method), in the most recent version of DemirciPRO, based on the past experience of the authors, only the pepper pot measurement design is provided. 
 
 \subsection{Simulation}
 
\begin{figure*}
 \centering
 \includegraphics[width=0.98\linewidth]{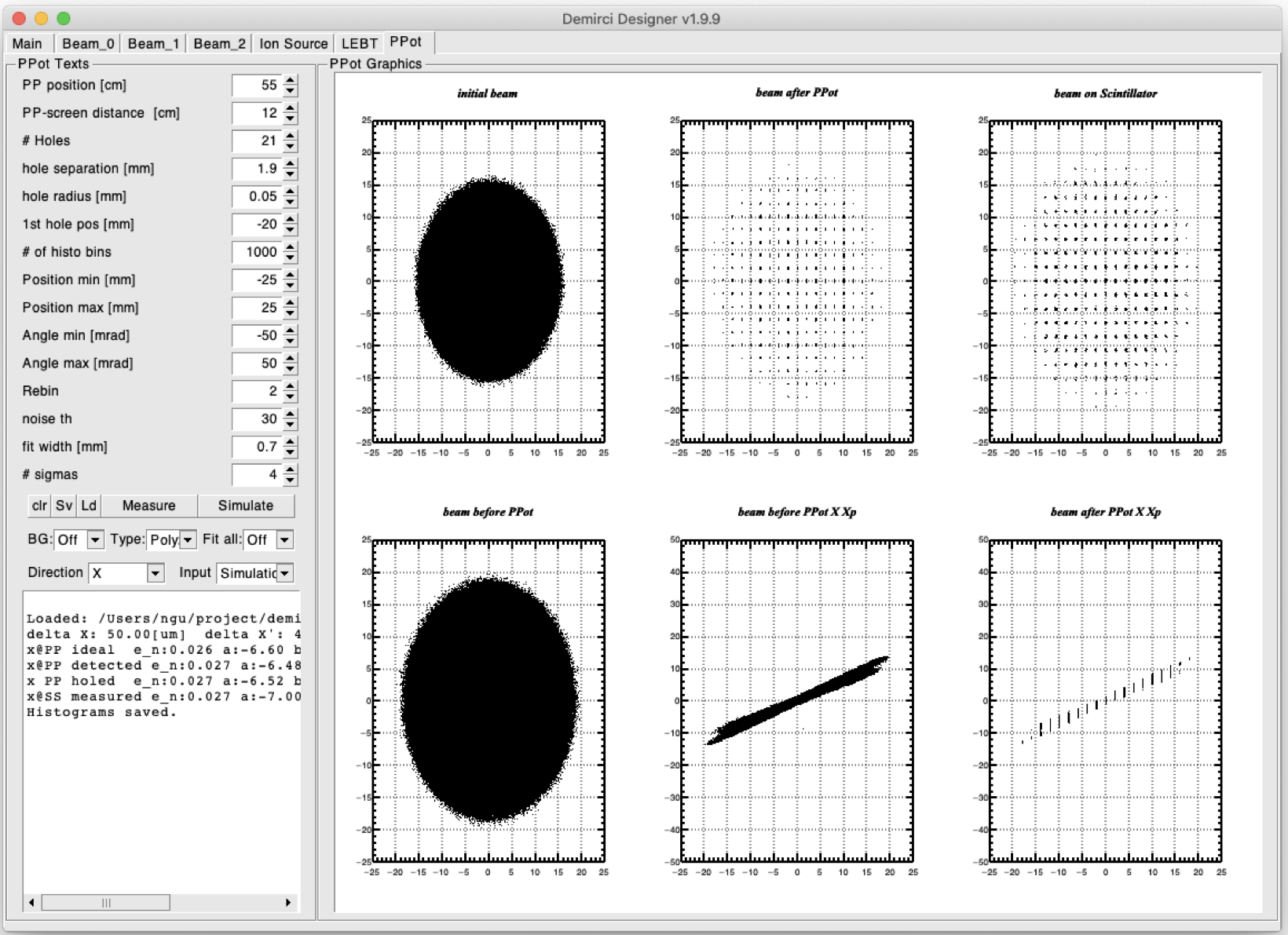}
 \caption{Pepper pot plate, simulation step output screen}
 \label{fig:ppplatesimu}
 \end{figure*}

 The design procedure starts by defining the position of the pepper pot plate (PPP) with respect to the beginning of the LEBT line. The second parameter to be defined is the distance between the PPP and the scintillating screen. These two, along with other geometrical parameters of the PPP (such as the number and radius of the holes) are to be entered using the left side of the design window similar to other tabs in DemirciPro. The GUI provides minimal help and default values for all parameters. Additionally, the whole measurement box setup, i.e. all relevant parameters can be saved and later reloaded using the appropriate GUI buttons, shown in Figure \ref{fig:ppplatesimu}. The next set of  parameters are related to the visualization of the simulated measurements: the bin counts and limit values of the histograms in physical and angular coordinates. 
 
 Once these are defined, the designer can simply simulate the proton beam going through the PPP and illuminating the scintillating screen. The results of such a simulation with 5 million events is shown in Figure \ref{fig:ppplatesimu}.
 When the simulation is finished, six plots (as a 2x3 matrix) are shown to the designer: 
The upper plot of the leftmost column  shows the initial beam as it is entering the measurement box, and the lower one the same beam right before hitting the PPP. The designer can check the enlargement of the beam spot using these two plots, as the PPP method is only suitable for divergent beams.
The protons surviving the PPP are shown as they are exiting the plate on the top row middle plot, and as they are hitting the scintillating screen on the same row, right plot. While the simulation is user selectable between X\&Y directions using the GUI, the program is set up to display and analyze only one of them at a time. Using the values from the simulation, the phase space of the beam, right before hitting the PPP is shown in the lower row middle plot whereas right after the PPP in the same row right plot. The plots are all calculated using truth information from the beam data and the resulting histograms are saved into a ROOT file for further analysis.
The text output section on the lower left side contains  some summary information related to the simulation. These are a number of emittance and Twiss parameter values calculated at different stages of the simulation. Those parameters are written as $\epsilon_n$, $\alpha$, $\beta$ (for normalized emittance, Twiss parameters alpha and beta) for brevity.
The "ideal" values are obtained by using all particles hitting the PPP; the "detected" values take into account the fact that the image detector (CCD camera) has a finite resolution due to its CCD pixel sizes. The CCD pixel size is obtained from the histogram parameters and it is represented by the value delta X. In example shown in Figure\ref{fig:ppplatesimu} it is 50~$\mu m$. From this point onward only the histogram bin center values are used as position information. The next set of values are called "holed" since only  particles which were able to pass through the holes of the PPP are used in these calculations. Finally, the "measured" values are calculated at the scintillator screen, i.e. after some designer defined drift distance.
 
 DemirciPro also updates the LEBT window to show the locations of the PPP and scintillator screen inside the measurement box.  The remaining part of the procedure is to analyze that data as if it were coming from a real measurement and to get the emittance values as close as possible to the ones calculated by the simulation part of the program. 
 
 \subsection{Analysis}
 The analysis of the image obtained from the scintillation screen  is performed using the two dimensional image obtained in the simulation stage. The algorithm analyzes the image as if the particle positions and angles were not known. This algorithm can easily be adapted to a real photo of the beamlets by converting the image to a 2D histogram. The analysis procedure consists of taking a projection of the 2D histogram along the X or Y directions to end up with a 1D histogram containing beamlet peaks. A peak finder (provided by ROOT library) determines the peak positions to fit a Gaussian function to each beamlet. The user can use the GUI to either fit each beamlet individually or all at once. If the number of simulated particles is not large enough, some regions of the scintillator screen receive less than ideal number of hits to create a reasonable distribution suitable to fitting. For this reason, it is possible to define a threshold value in DemirciPRO, below which the peak candidates are not considered. A possible cure to such a problem would be to rebin the 1D histogram, loose some details of the beamlet distributions but gain on the number of hits. This possibility is also provided via the GUI. During the fit, the initial values for the peak position and peak intensity are provided automatically by the peak finder function. The remaining parameter, the fit width can be defined via the DemirciPRO GUI. Once the fit is over, the beam emittance and other Twiss parameters can be calculated based on the available information. The program calculates the RMS emittance, however since the ion type and energy is known, it reports the normalized emittance value:
\begin{eqnarray}
\label{eq:epsi}
 \epsilon_{RMS} &=& \sqrt{\overline{x_h}^2\overline{x'}^2 -\overline{x_h x'}^2 }\\
 \epsilon_{n} &=& \beta \gamma  \epsilon_{RMS}  \quad , \nonumber 
\end{eqnarray}
where $\beta$ and $\gamma$ are relativistic functions, $x_h$ refers to the hole center positions in the selected direction (X or Y) and finally $x'$ is obtained from the analysis. Using the fact that the hole diameter is very small compared to the PPP to scintillator screen distance ($L$), the angle $x'$ can be approximated as:
\begin{equation}
 x' \equiv \frac{ f(h) -x_h}{L}    
\end{equation}
where $f(h)$ is the position of each bin center in the beamlet distribution. The counts in each bin, i.e. number of protons passing through the particular hole $h$ and hitting that bin can be obtained either from the actual data or using the Gaussian fit function.  The emittance value is estimated using the averages in equation \ref{eq:epsi} by summing on all holes or peaks above the user defined threshold value. 

\begin{figure*}
\centering
\includegraphics[width=0.98\linewidth]{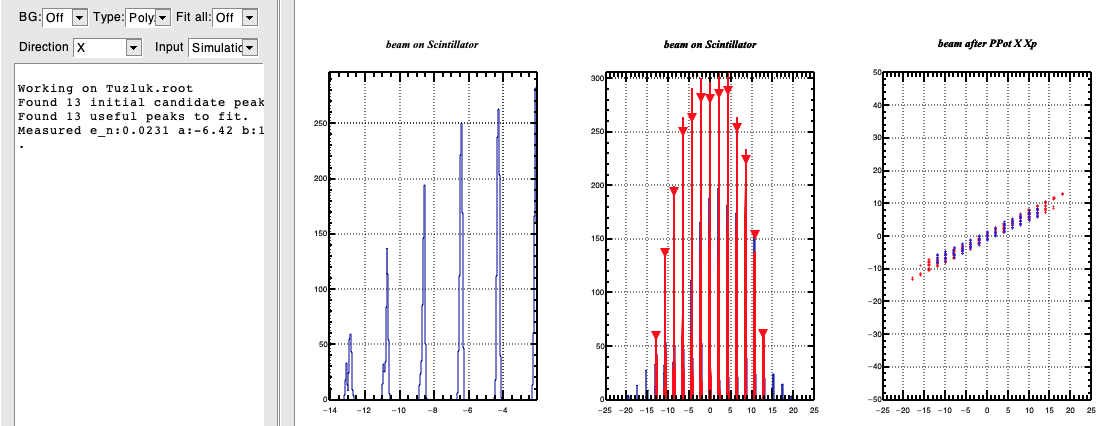}
\caption{Pepper pot plate measurement outputs, see text for details of the plots.
\label{fig:ppplatemeasure} }
 \end{figure*}

The output from a typical run is shown in Figure \ref{fig:ppplatemeasure}. Among the three plots, the left one shows the individual beamlet distributions zoomed to a particular section of the image, the central one shows the same beamlet intensity distributions together with the Gaussian fit functions overlaid in red, the red triangles show the beamlet intensity peak positions. The right graph shows the emittance plot. The truth information using the protons that survived the PPP is shown in red, whereas the measurement results using the scintillating screen are in blue. Note that the beamlet peaks below the pre-determined threshold are not taken into account, leaving some outlaying hole regions without corresponding measurement. The numerical results from this measurement is shown on the leftmost frame: $\epsilon_n = 0.0231~mm.mrad$. This value is to be compared to the truth information at various stages of the simulation as discussed in the previous section. 

\begin{table}
    \centering
    \caption{PPP emittance analysis results, the relative error is given with respect to the true value of 0.026~mm.mrad}
    \begin{tabular}{c|c|c}
    $\sigma$s   & $\epsilon_n ($mm.mrad) & \% error  \\ \hline
        4 &  0.0231  &  11.2 \\ 
        5 &  0.0247  & 5.0 \\ 
        8 &  0.0254  & 2.3 \\ 
        10 & 0.0254  & 2.3\\ 
    \end{tabular}
    \label{tab:PPP_results}
\end{table}

The effects of changing the number of sigmas of the Gaussian fits to consider, or other similar analysis parameters are to be explored by the PPP designer. Such an example study is presented in Table \ref{tab:PPP_results}, where the image from the simulation in the previous section is studied for an X-direction emittance measurement. Recalling its true value of 0.026~mm.mrad, the best relative error for this scenario is about~2\%.

 \section{Conclusions}
 \label{sec:conclusion}
 In the last three years, Demirci, the authors' GUI based  RFQ design software, has evolved into a GUI based light ion beamline designer software, called DemirciPro. The core code has been developed using C++ and it is based on ROOT libraries for graphical operations. For some of the modules, it relies on external libraries available freely under the GNU public license.  
 The ion source and the LEBT modules have been compared to a similar reference software for validation purposes. When the space charge effects are negligible, the relative difference between DemirciPRO and reference software modules is less than few percent. 
 
The previous version, namely Demirci, has been  used in the design of the SPP RFQ \cite{SPPRFQ}. Similarly, the current version, DemirciPro is being used in the design of a proton test beam, PTAK \cite{PTAK800}. As the software development continues, refinements such as the addition of space charge effects will be considered in the future. 

 \section{Acknowledgements}
 \label{sec:ack}
 This work is supported by TUBITAK project no. 117F143. The authors would like to thank Dr. S. Ogur for a careful reading of the manuscript.



\bibliographystyle{unsrt}
\bibliography{cas-refs}

\begin{thebibliography}{10}

\bibitem{Kapchinskii1969ALI}
I.~M. Kapchinskii and V.~A. Teplyakov.
\newblock A linear ion accelerator with spatially uniform hard focusing.
\newblock {\em Prib. Tekh. Eksp.}, 2:19--22, 1969.

\bibitem{Biscari1985ComputerPA}
C.~Biscari.
\newblock Computer programs and methods for the design of high intensity rfqs.
\newblock 1985.

\bibitem{Bondarev:2001in}
B.~Bondarev, A.~Durkin, Y.~Ivanov, I.~Shumakov, S.~Vinogradov, A.~Ovsyannikov,
  and D.~Ovsyannikov.
\newblock {The LIDOS.RFQ.Designer Development}.
\newblock {\em Conf. Proc. C}, 0106181:2947--2949, 2001.

\bibitem{Crandall1988PARMTEQABD}
Kenneth~R. Crandall and Thomas~P. Wangler.
\newblock Parmteq—a beam‐dynamics code fo the rfq linear accelerator.
\newblock volume 177, pages 22--28, 1988.

\bibitem{RFQGEN}
L.~M. Young and J.~Stoval.
\newblock Rfqgen user guide los alamos scientific lab.
\newblock 2017.

\bibitem{BENELOS}
R.~Duperrier.
\newblock High intensity beams dynamics in rfqs.
\newblock 2000.

\bibitem{Jang2006}
J.~H. Jang, Y.~S. Cho, K.~Y. Kim, Y.~H. Kim, and H.~J. Kwon.
\newblock {Beam dynamics of the PEFP Linac}.
\newblock volume 060626, pages 1612--1614, 2006.

\bibitem{ParmteqManual}
Kenneth~R. Crandall, Thomas~P. Wangler, Lloyd~M. Young, James~H. Billen,
  George~H. Neuschaefer, and Dale~L. Schrage.
\newblock {\em RFQ Design Codes}, 1986.

\bibitem{Pande_2015}
R.~Pande, P.~Singh, S.V.L.S. Rao, S.~Roy, and S.~Krishnagopal.
\newblock Optimization of solenoid based low energy beam transport line for
  high current h+ beams.
\newblock {\em Journal of Instrumentation}, 10(02):P02001--P02001, feb 2015.

\bibitem{Uriot2015}
D.~Uriot and N.~Pichoff.
\newblock {S}tatus of {T}race{W}in {C}ode.
\newblock In {\em Proc. 6th International Particle Accelerator Conference
  (IPAC'15), Richmond, VA, USA, May 3-8, 2015}, number~6 in International
  Particle Accelerator Conference, pages 92--94, Geneva, Switzerland, June
  2015. JACoW.

\bibitem{Stockli2006}
Martin~P. Stockli.
\newblock Measuring and analyzing the transverse emittance of charged particle
  beams.
\newblock {\em AIP Conference Proceedings}, 868(1):25--62, 2006.

\bibitem{Kalvas2010}
T.~Kalvas, O.~Tarvainen, T.~Ropponen, O.~Steczkiewicz, J.~Rje, and H.~Clark.
\newblock {IBSIMU: A three-dimensional simulation software for charged particle
  optics}.
\newblock {\em Review of Scientific Instruments}, 81(2):10--13, 2010.

\bibitem{Toivanen2013}
V.~Toivanen, T.~Kalvas, H.~Koivisto, J.~Komppula, and O.~Tarvainen.
\newblock {Double einzel lens extraction for the JYFL 14 GHz ECR ion source
  designed with IBSimu}.
\newblock {\em Journal of Instrumentation}, 8(5):1--19, 2013.

\bibitem{Brun:1997pa}
R.~Brun and F.~Rademakers.
\newblock {ROOT: An object oriented data analysis framework}.
\newblock {\em Nucl. Instrum. Meth. A}, 389:81--86, 1997.

\bibitem{IBSimuManual}
T.~Kalvas.
\newblock {\em {Reference manual for Ion Beam Simulator 1.0.6}}, 2015.

\bibitem{TRAVELManual}
J.F. Perrin, A.and~Amand, J.B. Mutze, T.and~Lallement, and S.~Lanzone.
\newblock {\em {TRAVEL v4.07 User Manual, CERN}}, 2007.

\bibitem{PlotwinManual}
D.~Uriot.
\newblock {\em {Plotwin v4.0 User Manual, CEA Saclay}}, 2017.

\bibitem{Keller1999}
R.~Keller.
\newblock {Ion-source and low-energy beam-transport issues for H-
  accelerators}.
\newblock {\em Proceedings of the IEEE Particle Accelerator Conference},
  1:87--91, 1999.

\bibitem{Prost2016}
Lionel Prost.
\newblock {Selected List of Low Energy Beam Transport Facilities for Light-Ion,
  High-Intensity Accelerators}.
\newblock {\em arXiv preprint arXiv:1602.05488}, pages 1--7, 2016.

\bibitem{Chauvin2011}
N.~Chauvin, O.~Delferri{\`{e}}re, R.~Duperrier, R.~Gobin, P.~A.P. Nghiem, and
  D.~Uriot.
\newblock {Source and injector design for intense light ion beams including
  space charge neutralisation}.
\newblock {\em Proceedings - 25th Linear Accelerator Conference, LINAC 2010},
  pages 740--744, 2011.

\bibitem{winehq}
H.~et~al. Davies.
\newblock {\em {Reference manual for Wine}}, 2021.

\bibitem{SPPRFQ}
B.~Yasatekin, G.~Turemen, A.~Alacakir, and G.~Unel.
\newblock Design studies with demirci for spp rfq, 2015.

\bibitem{PTAK800}
A.~Adiguzel et~al.
\newblock Design and construction of a proton testbeam at kahvelab.
\newblock (In preparation), 2021.

\end{thebibliography}
\end{document}